\documentclass{iaus}
\usepackage{graphicx}

\title[Search for the magnetic field of $\xi$ Persei O7.5 III] %% give here short title %%
{Search for the magnetic field of the O7.5 III star $\xi$ Persei}

\author[Henrichs, Schnerr, de Jong, Kaper, Donati,  Catala]   %% give here short author list %%
{H.F. Henrichs$^1$%
,
R.S. Schnerr$^2$, J.A. de Jong$^3$, L. Kaper$^1$, J.-F.~Donati$^4$
\and C. Catala$^5$}
\affiliation{$^1$Astronomical Institute, University of Amsterdam, Amsterdam, Netherlands
\\[\affilskip]
$^2$Inst.\ for Solar Physics, Royal Swedish Academy of Sciences,
Stockholm, Sweden
\\[\affilskip]
$^3$Max Planck Institute for Extraterrestrial Physics, Garching, Germany 
\\[\affilskip]
$^4$4Observatoire Midi-Pyr\'{e}n\'{e}es, Toulouse, France,
$^5$LESIA, Observatoire de Paris, CNRS, Universit\'{e} Paris Diderot; 5 place Jules Janssen, 92190 Meudon, France
}
\pubyear{2009}
\volume{259}  %% insert here IAU Symposium No.
\pagerange{100--100}
\date{November 2008 and in revised form ??}
\jname{Cosmic Magnetic Fields: From Planets, to Stars and Galaxies}
\editors{K.G. Strassmeier, A.G. Kosovichev \& J.E. Beckman, eds.}
\graphicspath{{converted_graphics/}}
\begin{document}

\maketitle

\begin{abstract}
Cyclical wind variability is an ubiquitous but as yet unexplained feature among OB stars. The O7.5 III(n)((f)) star $\xi$ Persei is the brightest representative of this class on the Northern hemisphere. As its prominent cyclical wind properties vary on a rotational time scale (2 or 4 days) the star has been already for a long time a serious magnetic candidate. As the cause of this enigmatic behavior non-radial pulsations and/or a surface magnetic field  are suggested. We present a preliminary report on our attempts to detect a magnetic field in this star with high-resolution measurements obtained with the spectropolarimeter Narval at TBL, France during 2 observing runs of 5 nights in 2006 and 5 nights in 2007. Only upper limits could be obtained, even with the longest possible exposure times. If the star hosts a magnetic field, its surface strength should be less than about 300 G. This would still be enough to disturb the stellar wind significantly. From our new data it seems that the amplitude of the known non-radial pulsations has changed within less than a year, which needs further investigation.

\end{abstract}

\section{Introduction}
Like many O and B stars, the O7.5III(n)((f)) star $\xi$ Per shows very prominent cyclical wind variability in the UV resonance lines (Fig.~1a), manifested by discrete absorption components (DACs) which migrate from red to blue, and narrow when they approach (but not reach) the terminal velocity as measured from saturated wind profiles (\textit{e.g.} \cite[Kaper \etal\ 1999)]{kaper99}. In $\xi$ Per the DAC period is 2.09 d.  Multiwavelength observations of a number of OB stars, including $\xi$ Per, have shown that the cyclic behavior is present down to the surface of the star (\cite[de Jong \etal\ 1997)]{dejong97}, and that the typical timescale varies with the (estimated) rotational timescale. This strongly argues in favor of a surface phenomenon which perturbs the base of the flow.  Two scenarios are proposed: in the so-called Corotating Interaction Region (CIR) model a perturbation at the surface of the star causes a local increase (or decrease) of the radiative force driving the stellar wind, or surface magnetic fields may disturb the outflow, both resulting in a rotationally modulated stellar wind (see \cite[Cranmer and Owocki 1996)]{cranmer96}. A number of coordinated UV and optical observations have confirmed the CIR model, including for the case of $\xi$ Per, but the origin of the perturbations is not known, which is one of the most challenging problems in stellar wind research of the last decades. 
We present here our most recent efforts to measure the magnetic field and the pulsation properties.

\begin{figure}[htp]
\centering
\leftline{\includegraphics[bb=0 13 596 842,height=4.5cm,keepaspectratio]{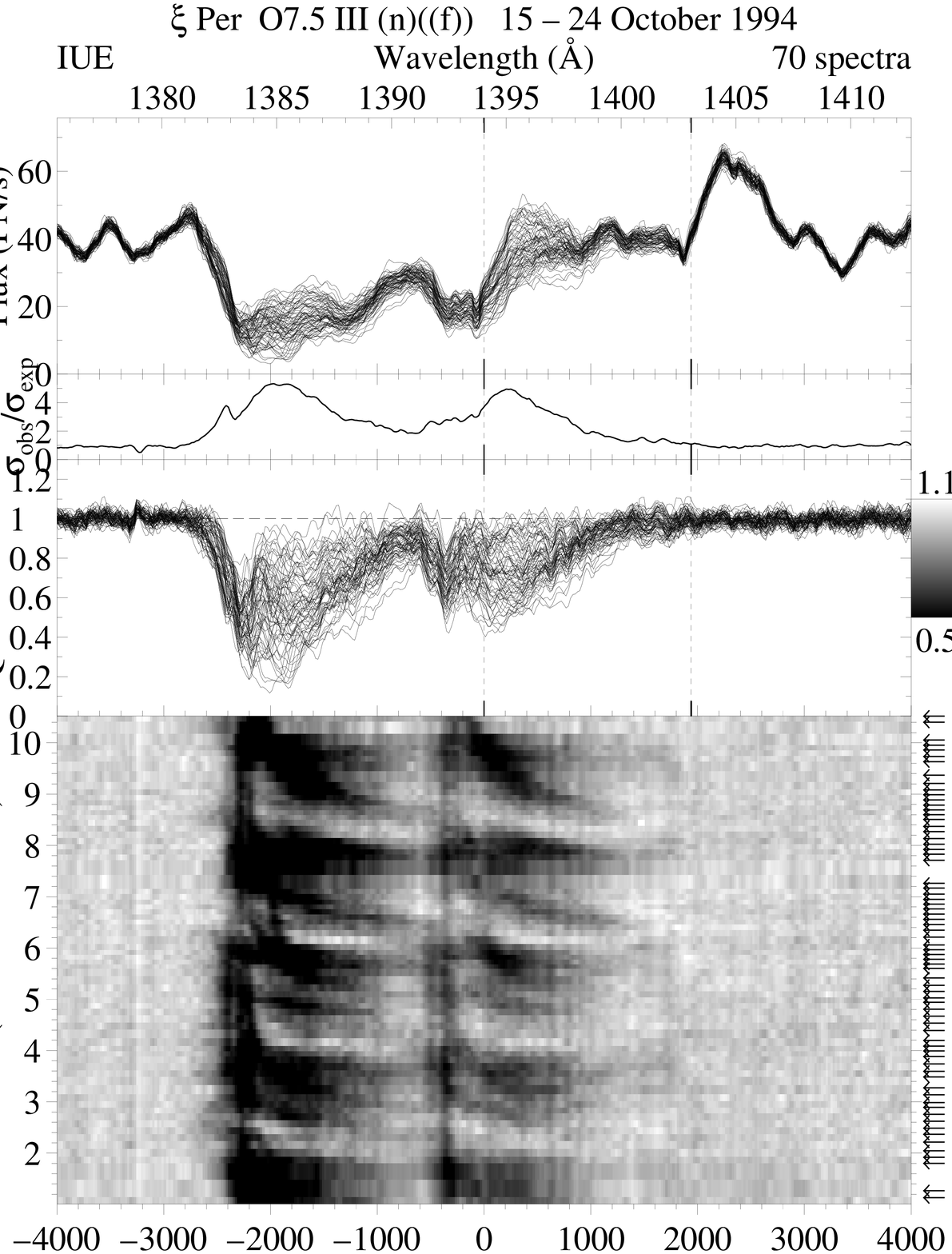}}

\vspace{-4.7cm}
\centerline{\includegraphics[bb=38 218 551 770,height=4.5cm,keepaspectratio]{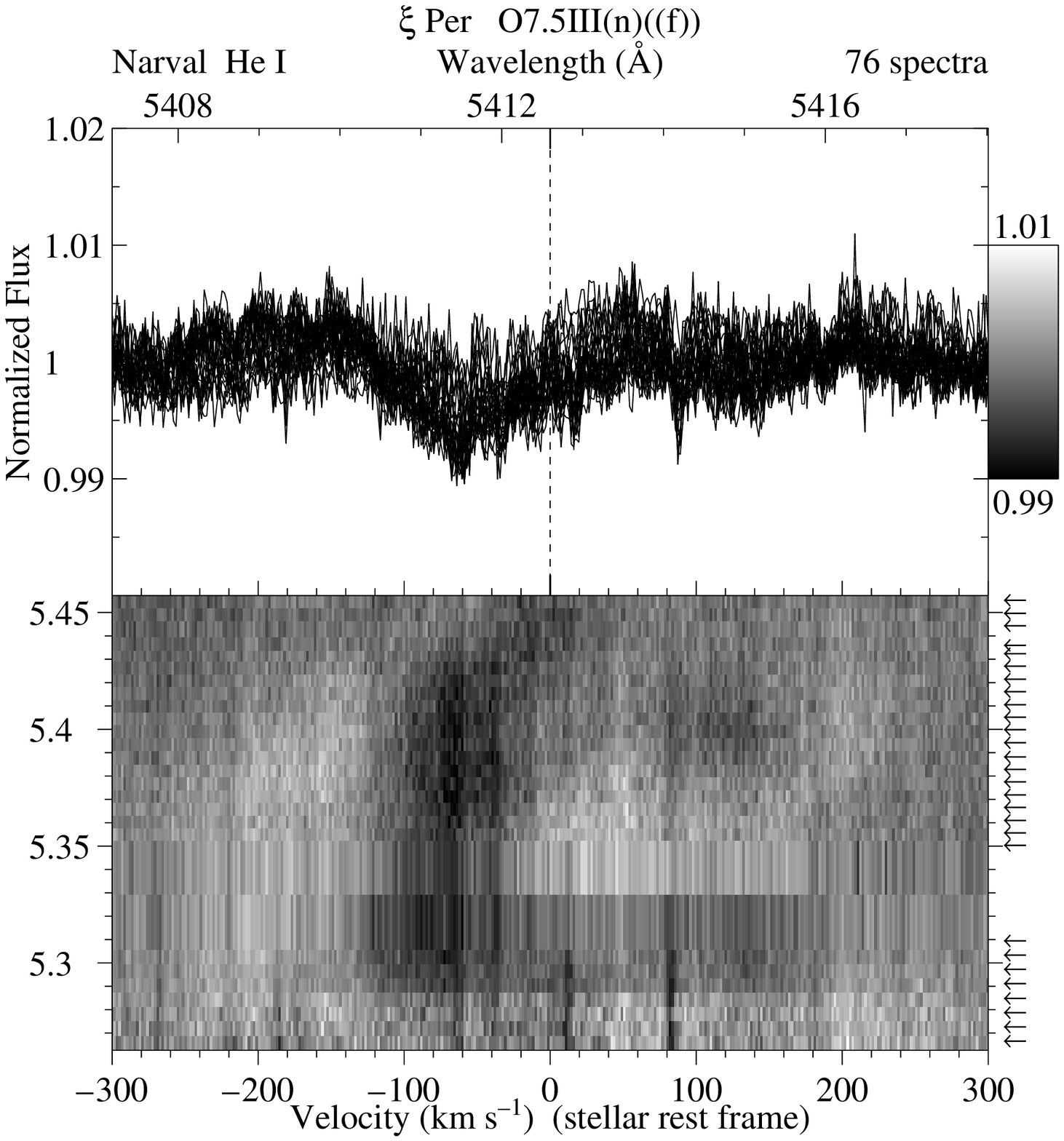}}

\vspace{-4.6cm}
\rightline{\includegraphics[bb=38 203 551 770,height=4.7cm,keepaspectratio]{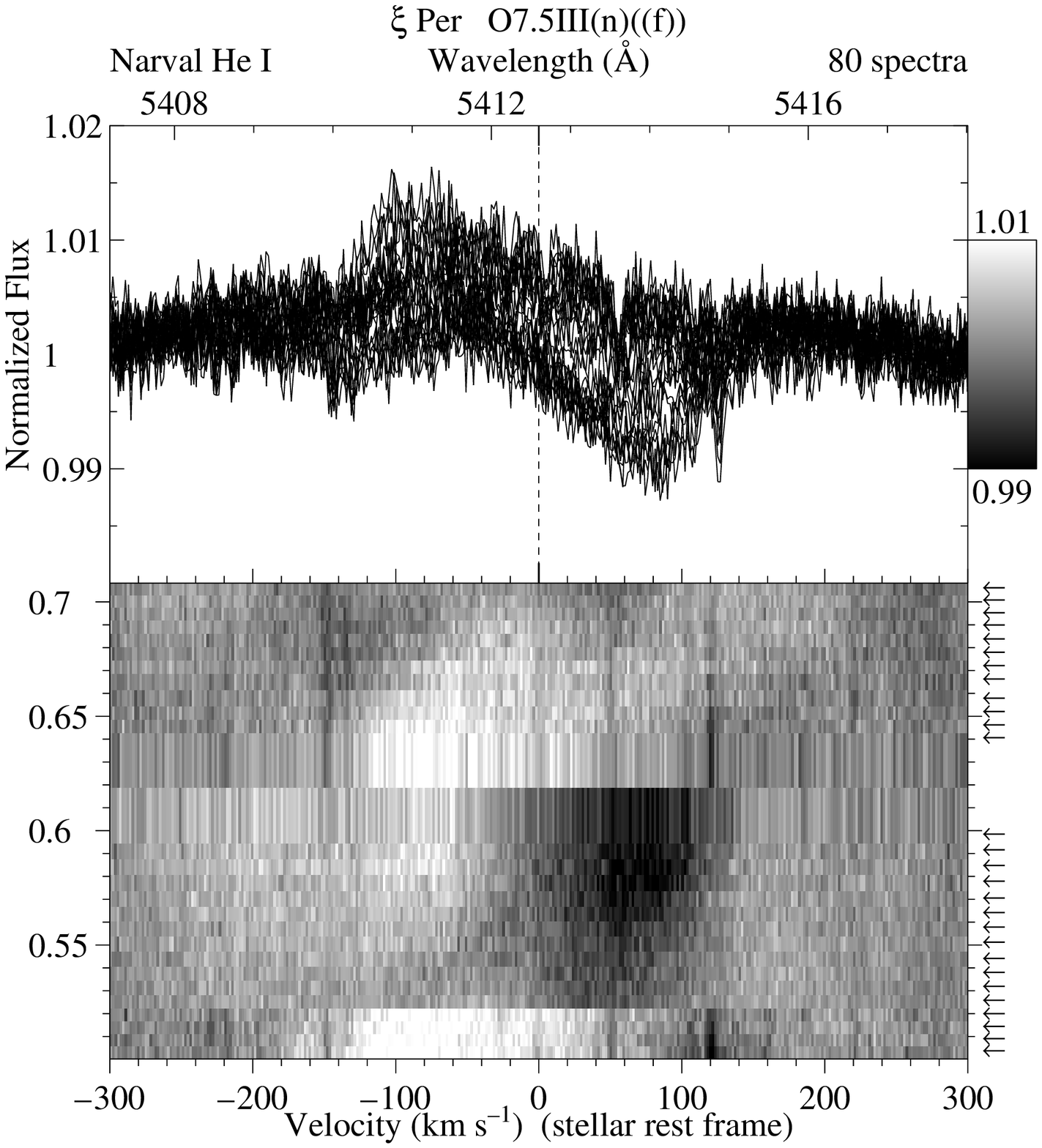}}

\vspace{0.1cm}
\caption{(a) Progressing DACs in the Si IV wind lines in 1994, (b) Non radial pulsation (period 3.5 h) in the He I 5411 line, December 2006.
 (c)  September 2007; the amplitude of the moving features seems to have changed after 9 months.
}
\end{figure}

\begin{figure}[htp]
\centering
\leftline{\includegraphics[bb=30 378 507 742,height=3.2cm,keepaspectratio]{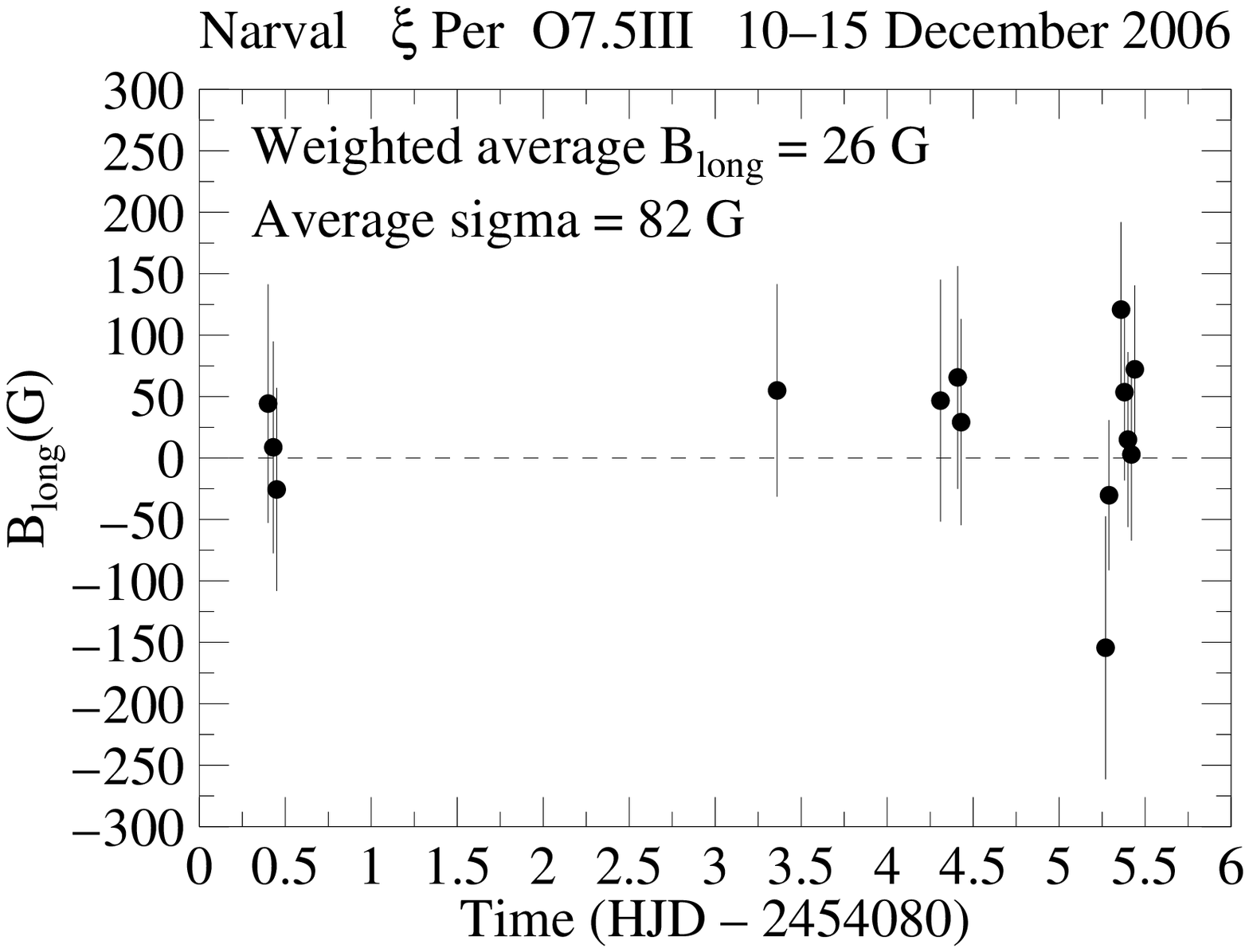}}
\vspace{-3.2cm}
\centerline{\includegraphics[bb=30 378 507 742,height=3.2cm,keepaspectratio]{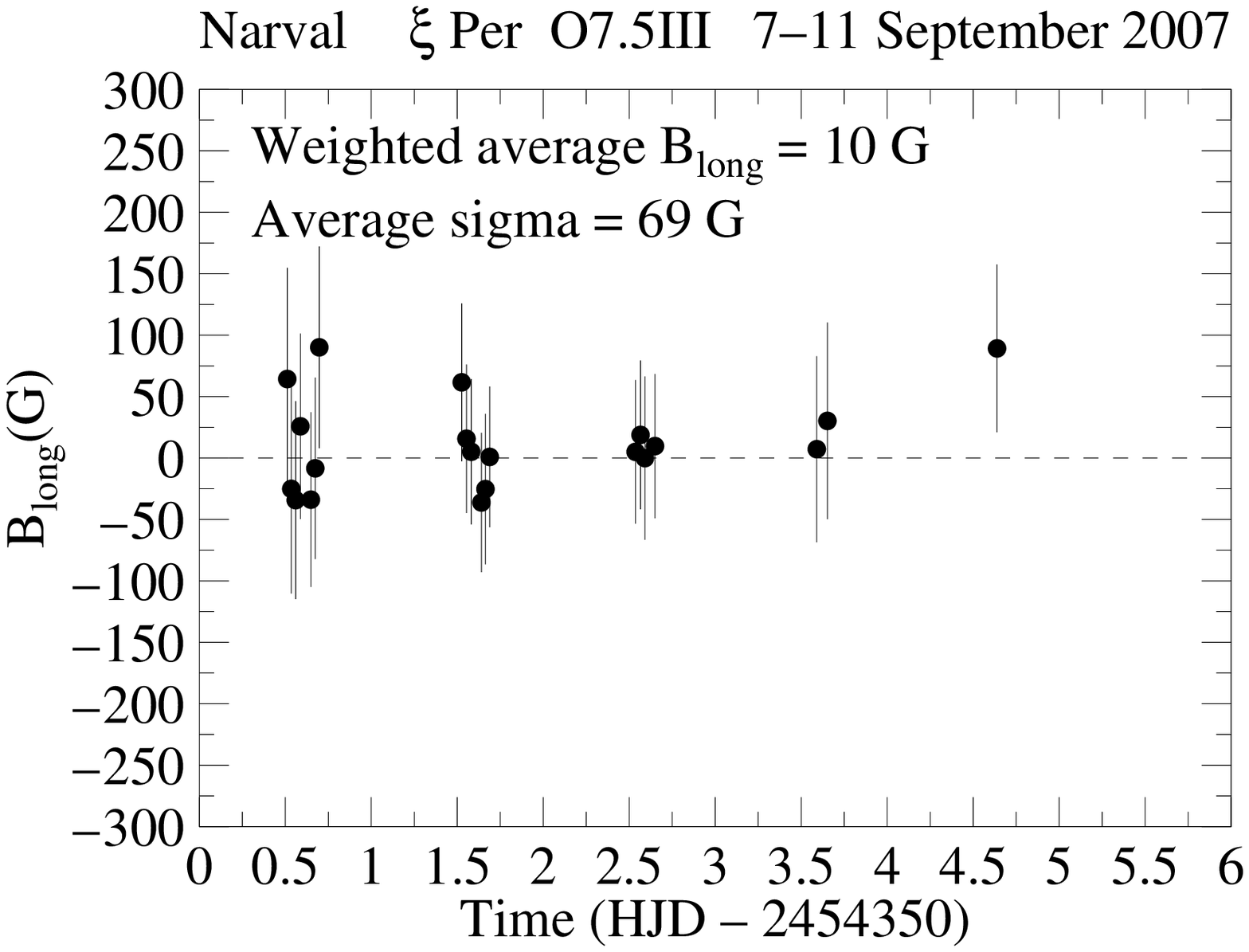}}
\vspace{-3.2cm}
\rightline{\includegraphics[bb=8 240 503 632,height=3.2cm,keepaspectratio]{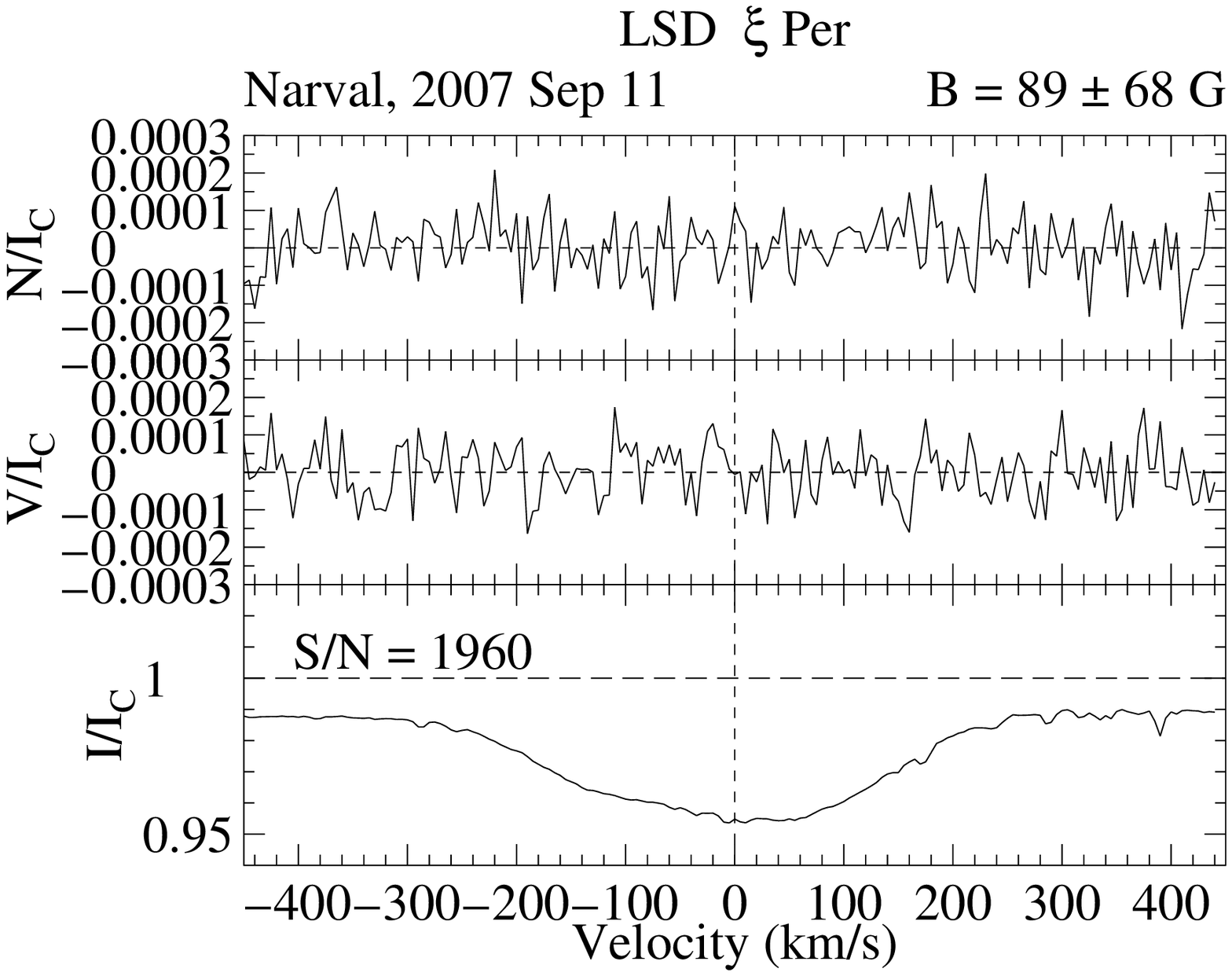}}

\caption{Preliminary magnetic results of Narval spectropolarimetry as a function of time. (a) December 2006. (b) September 2007. (c) LSD Stokes V profile of the last point in September 2007.  No significant Zeeman signature was found.
}
\end{figure}

\section{Data analysis and discussion}
Previous Musicos magnetic measurements of $\xi$ Per were presented by \cite[de Jong \etal\ (2001)]{dejong01}, with no detection. For Narval data the magnetic analysis is essentially the same, applying the least-squares deconvolution method (\cite[Donati \etal\ 1997)]{donati97} to the spectral lines sensitive to magnetic effects. This yields the longitudinal component of the field, averaged over the facing hemisphere of the star. In Fig.~2a and 2b the 45 results are plotted as a function of time. No Zeeman signature was found. As an example, Fig.~2c shows the LSD profile of the best exposed spectrum of September 2007 with a S/N = 1960. The magnetic values are preliminary, as the used spectral linelist can still be optimized.
 
We also analysed a number of spectral lines for non-radial pulsations with known period of 3.5 h found in a previous Musicos campaign (\cite[de Jong \etal\ 1999]{dejong99}). The amplitude may have changed during the 9 months between our Narval runs, see Fig.~1b and 1c. This obviously needs further investigation, as the impact on the stellar wind may have changed. 

Observations with a higher efficiency should be able to detect a possible field, unless the geometry is such that the magnetic properties of the two hemispheres cancel out.

\end{document}